\input harvmac.tex

\def\inbar{\vrule height1.5ex width.4pt depth0pt}

\def\IN{\relax{\rm I\kern-.18em N}}
\def\IB{\relax\hbox{$\inbar\kern-.3em{\rm B}$}}
\def\IC{\relax\hbox{$\inbar\kern-.3em{\rm C}$}}
\def\IQ{\relax\hbox{$\inbar\kern-.3em{\rm Q}$}}
\def\ID{\relax\hbox{$\inbar\kern-.3em{\rm D}$}}
\def\IE{\relax\hbox{$\inbar\kern-.3em{\rm E}$}}
\def\IF{\relax\hbox{$\inbar\kern-.3em{\rm F}$}}
\def\IG{\relax\hbox{$\inbar\kern-.3em{\rm G}$}}
\def\IGa{\relax\hbox{${\rm I}\kern-.18em\Gamma$}}
\def\IH{\relax{\rm I\kern-.18em H}}
\def\IK{\relax{\rm I\kern-.18em K}}
\def\IL{\relax{\rm I\kern-.18em L}}
\def\IP{\relax{\rm I\kern-.18em P}}
\def\IR{\relax{\rm I\kern-.18em R}}

\def\Z{\relax\ifmmode\mathchoice{\hbox{\cmss Z\kern-.4em Z}}{\hbox{\cmss Z\kern-.4em Z}} {\lower.9pt\hbox{\cmsss Z\kern-.4em Z}}{\lower1.2pt\hbox{\cmsss Z\kern-.4em Z}}\else{\cmss Z\kern-.4em Z}\fi}

\def\II{\relax{\rm I\kern-.18em I}}
\def\one{\relax{\rm 1\kern-.25em I}}

\def\Z{{\bf Z}}

\def\CLL{\relax{\CL\kern-.74em \CL}}

\def\Kah{K\"ahler}
\def\Poin{Poincar\'e}

\def\del{\partial}

\def\inv{^{-1}}

\def\half{\textstyle{1\over2}}
\def\third{\textstyle{1\over3}}
\def\quart{\textstyle{1\over4}}
\def\fifth{\textstyle{1\over5}}

\def\cC{{\cal C}}

\def\cG{{\cal G}}
\def\cH{{\cal H}}

\def\cL{{\cal L}}

\def\cN{{\cal N}}
\def\cO{{\cal O}}
\def\cP{{\cal P}}

\def\cW{{\cal W}}

\def\CC{{\cal C}}

\def\CL{{\cal L}}

\def\a{\alpha}
\def\b{\beta}

\def\th{\theta}

\noblackbox

\parskip=4pt plus 15pt minus 1pt
\baselineskip=15pt plus 2pt minus 1pt

 at 13truept
 at 13truept
 at 13truept

\def\ts{\textstyle}

\def\tablecaption#1#2{\kern.75truein\lower0truept\hbox{\vbox{\hsize=5truein\noindent{\bf Table\hskip5truept#1:} #2}}}

\def\-{\hphantom{-}}

\def\({\bigl(}
\def\){\bigr)}
\def\<{\langle\,}
\def\>{\,\rangle}

\def\abstract#1{\vskip .5in\vfil\centerline{\bf Abstract}\penalty1000{{\smallskip\ifx\answ\bigans\leftskip 2pc \rightskip 2pc\else\leftskip 5pc \rightskip 5pc\fi\noindent\abstractfont \baselineskip=12pt{#1} \smallskip}}\penalty-1000}

\def\abstractmod#1{\vskip.5in\vfil\centerline{\bf Abstract}\penalty1000{{\smallskip\leftskip0pc\rightskip0pc\noindent\abstractfont\baselineskip=12.5pt{#1}\smallskip}}\penalty-1000}

\def\us#1{\underline{#1}}
\def\hth/#1#2#3#4#5#6#7{{\tt hep-th/#1#2#3#4#5#6#7}}
\def\nup#1({Nucl.\ Phys.\ $\us {B#1}$\ (}
\def\plt#1({Phys.\ Lett.\ $\us {B#1}$\ (}
\def\cmp#1({Comm.\ Math.\ Phys.\ $\us {#1}$\ (}
\def\prp#1({Phys.\ Rep.\ $\us {#1}$\ (}
\def\prl#1({Phys.\ Rev.\ Lett.\ $\us {#1}$\ (}
\def\prv#1({Phys.\ Rev.\ $\us {#1}$\ (}
\def\mpl#1({Mod.\ Phys.\ Let.\ $\us {A#1}$\ (}
\def\ijmp#1({Int.\ J.\ Mod.\ Phys.\ $\us{A#1}$\ (}

\def\mn{\the\secno.\the\subsecno}
\def\br{\hfill\break}
\def\noi{\noindent}

\def\fermat#1#2#3#4#5{x_1^{#1}+x_2^{#2}+x_3^{#3}+x_4^{#4}+x_5^{#5}}
\def\x#1{x_{#1}}
\def\hphi{\hat{\phi}}

\lref\MW{
  D.~R.~Morrison and J.~Walcher,
  ``D-branes and Normal Functions,''
  arXiv:0709.4028 [hep-th].
}
\lref\KW{
  D.~Krefl and J.~Walcher,
  ``Real Mirror Symmetry for One-parameter Hypersurfaces,''
  JHEP {\bf 0809}, 031 (2008)
  [arXiv:0805.0792 [hep-th]].
}
\lref\Knapp{
  J.~Knapp and E.~Scheidegger,
  ``Towards Open String Mirror Symmetry for One-Parameter Calabi-Yau
  Hypersurfaces,''
  arXiv:0805.1013 [hep-th].
}
\lref\JS{
  H.~Jockers and M.~Soroush,
  ``Effective superpotentials for compact D5-brane Calabi-Yau geometries,''
  Commun.\ Math.\ Phys.\  {\bf 290}, 249 (2009)
  [arXiv:0808.0761 [hep-th]].
}
\lref\Mayr{
  M.~Alim, M.~Hecht, P.~Mayr and A.~Mertens,
  ``Mirror Symmetry for Toric Branes on Compact Hypersurfaces,''
  arXiv:0901.2937 [hep-th].
}
\lref\JockersMN{
  H.~Jockers and M.~Soroush,
  ``Relative periods and open-string integer invariants for a compact
  Calabi-Yau hypersurface,''
  arXiv:0904.4674 [hep-th].
}
\lref\LercheCK{
  W.~Lerche, P.~Mayr and N.~Warner,
  ``Holomorphic N = 1 special geometry of open-closed type II strings,''
  arXiv:hep-th/0207259.
}
\lref\LercheYW{
  W.~Lerche, P.~Mayr and N.~Warner,
  ``N = 1 special geometry, mixed Hodge variations and toric geometry,''
  arXiv:hep-th/0208039.
}
\lref\MayrZI{
  P.~Mayr,
  ``Summing up open string instantons and N = 1 string amplitudes,''
  arXiv:hep-th/0203237.
}
\lref\LercheCW{
  W.~Lerche and P.~Mayr,
  ``On N = 1 mirror symmetry for open type II strings,''
  arXiv:hep-th/0111113.
}
\lref\Mayrone{
  P.~Mayr,
  ``N = 1 mirror symmetry and open/closed string duality,''
  Adv.\ Theor.\ Math.\ Phys.\  {\bf 5}, 213 (2002)
  [arXiv:hep-th/0108229].
}

\lref\KMP{
  S.~H.~Katz, D.~R.~Morrison and M.~Ronen~Plesser,
  ``Enhanced Gauge Symmetry in Type II String Theory,''
  Nucl.\ Phys.\  B {\bf 477}, 105 (1996)
  [arXiv:hep-th/9601108].
}
\lref\lerche{
  W.~Lerche,
  ``Fayet-Iliopoulos potentials from four-folds,''
  JHEP {\bf 9711}, 004 (1997)
  [arXiv:hep-th/9709146].
}
\lref\GVW{
  S.~Gukov, C.~Vafa and E.~Witten,
  ``CFT's from Calabi-Yau four-folds,''
  Nucl.\ Phys.\  B {\bf 584}, 69 (2000)
  [Erratum-ibid.\  B {\bf 608}, 477 (2001)]
  [arXiv:hep-th/9906070].
}
\lref\MayrSH{
  P.~Mayr,
  ``Mirror symmetry, N = 1 superpotentials and tensionless strings on
  Calabi-Yau four-folds,''
  Nucl.\ Phys.\  B {\bf 494}, 489 (1997)
  [arXiv:hep-th/9610162].
}
\lref\GukovIQ{
  S.~Gukov and M.~Haack,
  ``IIA string theory on Calabi-Yau fourfolds with background fluxes,''
  Nucl.\ Phys.\  B {\bf 639}, 95 (2002)
  [arXiv:hep-th/0203267].
}
\lref\GV{
  R.~Gopakumar and C.~Vafa,
  ``M-theory and topological strings. I,''
  arXiv:hep-th/9809187.
}
\lref\OV{
  H.~Ooguri and C.~Vafa,
  ``Knot invariants and topological strings,''
  Nucl.\ Phys.\  B {\bf 577}, 419 (2000)
  [arXiv:hep-th/9912123].
}
\lref\phases{
  E.~Witten,
  ``Phases of N = 2 theories in two dimensions,''
  Nucl.\ Phys.\  B {\bf 403}, 159 (1993)
  [arXiv:hep-th/9301042].
}
\lref\D{
  M.~Bershadsky, C.~Vafa and V.~Sadov,
  ``D-Strings on D-Manifolds,''
  Nucl.\ Phys.\  B {\bf 463}, 398 (1996)
  [arXiv:hep-th/9510225].
}
\lref\AV{
  M.~Aganagic and C.~Vafa,
  ``Mirror symmetry, D-branes and counting holomorphic discs,''
  arXiv:hep-th/0012041.
}
\lref\AKV{
  M.~Aganagic, A.~Klemm and C.~Vafa,
  ``Disk instantons, mirror symmetry and the duality web,''
  Z.\ Naturforsch.\  A {\bf 57}, 1 (2002)
  [arXiv:hep-th/0105045].
}
\lref\AK{
  M.~Aganagic, A.~Karch, D.~Lust and A.~Miemiec,
  ``Mirror symmetries for brane configurations and branes at singularities,''
  Nucl.\ Phys.\  B {\bf 569}, 277 (2000)
  [arXiv:hep-th/9903093].
}
\lref\SYZ{
  A.~Strominger, S.~T.~Yau and E.~Zaslow,
  ``Mirror symmetry is T-duality,''
  Nucl.\ Phys.\  B {\bf 479}, 243 (1996)
  [arXiv:hep-th/9606040].
}
\lref\Walcher{
  J.~Walcher,
  ``Opening mirror symmetry on the quintic,''
  Commun.\ Math.\ Phys.\  {\bf 276}, 671 (2007)
  [arXiv:hep-th/0605162].
}
\lref\gaberdiel{
  M.~Baumgartl, I.~Brunner and M.~R.~Gaberdiel,
  ``D-brane superpotentials and RG flows on the quintic,''
  JHEP {\bf 0707}, 061 (2007)
  [arXiv:0704.2666 [hep-th]].
}
\lref\DenefWQ{
  F.~Denef,
  ``Les Houches Lectures on Constructing String Vacua,''
  arXiv:0803.1194 [hep-th].
}
\lref\TV{
  M.~Aganagic, A.~Klemm, M.~Marino and C.~Vafa,
  ``The topological vertex,''
  Commun.\ Math.\ Phys.\  {\bf 254}, 425 (2005)
  [arXiv:hep-th/0305132].
}
\lref\WalcherUJ{
  J.~Walcher,
  ``Calculations for Mirror Symmetry with D-branes,''
  arXiv:0904.4905 [hep-th].
}
\lref\KL{
  S.~Katz and M.~Liu,
  ``Enumerative geometry of stable maps with Lagrangian boundary conditions
  and multiple covers of the disc,''
  Adv.Theor.Math.Phys. 5 (2002) 1-49 and Geom. Topol. Monogr. 8 (2006) 1-47.
}
\lref\MNOP{
  D.~Maulik, N.~Nekrasov, A.~Okounkov and R.~Pandharipande,
  ``Gromov-Witten theory and Donaldson-Thomas theory. I, II,''
  arXiv:math/0312059; arXiv:math/0406092.\br
  D.~Maulik, A.~Oblomkov, A.~Okounkov and R.~Pandharipande,
  ``Gromov-Witten/ Donaldson-Thomas correspondence for toric threefolds,''
  arXiv:0809.3976 [math.AG].
}
\lref\Griffiths{
  P.~A.~Griffiths,
  ``On the periods of certain rational integrals. I, II,''
  Ann. of Math. (2) 90 (1969), 460Ð495; ibid. (2) 90 (1969) 496Ð541.\br
  P.~A.~Griffiths,
  ``A theorem concerning the differential equations satisfied by normal
  functions associated to algebraic cycles,''
  Amer. J. Math. 101 (1979) 94Ð131.
}
\lref\Shamit{
  S.~Kachru, S.~H.~Katz, A.~E.~Lawrence and J.~McGreevy,
  ``Open string instantons and superpotentials,''
  Phys.\ Rev.\  D {\bf 62}, 026001 (2000)
  [arXiv:hep-th/9912151].
}
\lref\PSW{
  R.~Pandharipande, J.~Solomon and J.~Walcher,
  ``Disk enumeration on the quintic 3-fold,''
  arXiv:math/0610901v2.
}
\lref\MM{
  M.~Aganagic, A.~Klemm, M.~Marino and C.~Vafa,
  ``Matrix model as a mirror of Chern-Simons theory,''
  JHEP {\bf 0402}, 010 (2004)
  [arXiv:hep-th/0211098].
}
\lref\ShamitB{
  S.~Kachru, S.~H.~Katz, A.~E.~Lawrence and J.~McGreevy,
  ``Mirror symmetry for open strings,''
  Phys.\ Rev.\  D {\bf 62}, 126005 (2000)
  [arXiv:hep-th/0006047].
}
\lref\GomisWC{
  J.~Gomis, F.~Marchesano and D.~Mateos,
  ``An open string landscape,''
  JHEP {\bf 0511}, 021 (2005)
  [arXiv:hep-th/0506179].
}
\lref\old{
  M. Aganagic and C.Vafa, Unpublished.
}
\lref\MarinoAF{
  M.~Marino, R.~Minasian, G.~W.~Moore and A.~Strominger,
  ``Nonlinear instantons from supersymmetric p-branes,''
  JHEP {\bf 0001}, 005 (2000)
  [arXiv:hep-th/9911206].
}
\lref\CollinucciPF{
  A.~Collinucci, F.~Denef and M.~Esole,
  ``D-brane Deconstructions in IIB Orientifolds,''
  JHEP {\bf 0902}, 005 (2009)
  [arXiv:0805.1573 [hep-th]].
}
\lref\AVG{
  M.~Aganagic and C.~Vafa,
  ``G(2) manifolds, mirror symmetry and geometric engineering,''
  arXiv:hep-th/0110171.
}
\lref\DenefVG{
  F.~Denef and G.~W.~Moore,
  ``Split states, entropy enigmas, holes and halos,''
  arXiv:hep-th/0702146.
}
\lref\GMP{
  B.~R.~Greene, D.~R.~Morrison and M.~R.~Plesser,
  ``Mirror manifolds in higher dimension,''
  Commun.\ Math.\ Phys.\  {\bf 173}, 559 (1995)
  [arXiv:hep-th/9402119].
}
\lref\HuangHQ{
  M.~x.~Huang, A.~Klemm and S.~Quackenbush,
  ``Topological String Theory on Compact Calabi-Yau: Modularity and Boundary
  Conditions,''
  Lect.\ Notes Phys.\  {\bf 757}, 45 (2009)
  [arXiv:hep-th/0612125].
}
\lref\GrimmDQ{
  T.~W.~Grimm, T.~W.~Ha, A.~Klemm and D.~Klevers,
  ``The D5-brane effective action and superpotential in N=1
  compactifications,''
  Nucl.\ Phys.\  B {\bf 816}, 139 (2009)
  [arXiv:0811.2996 [hep-th]].
}
\lref\AlimBX{
  M.~Alim, M.~Hecht, H.~Jockers, P.~Mayr, A.~Mertens and M.~Soroush,
  ``Hints for Off-Shell Mirror Symmetry in type II/F-theory
  Compactifications,''
  arXiv:0909.1842 [hep-th].
}
\lref\GrimmEF{
  T.~W.~Grimm, T.~W.~Ha, A.~Klemm and D.~Klevers,
  ``Computing Brane and Flux Superpotentials in F-theory Compactifications,''
  arXiv:0909.2025 [hep-th].
}
\lref\MartucciIJ{
  L.~Martucci,
  ``D-branes on general N = 1 backgrounds: Superpotentials and D-terms,''
  JHEP {\bf 0606}, 033 (2006)
  [arXiv:hep-th/0602129].
}
\lref\HV{
  K.~Hori and C.~Vafa,
  ``Mirror symmetry,''
  arXiv:hep-th/0002222.
}
\nopagenumbers
\pageno=0

{\Title{\vbox{\hbox{}}}{\vbox{
\centerline{The Geometry of D-Brane Superpotentials}
}}

\vskip1cm
\centerline{\bf Mina Aganagic${}^{a}$ and Christopher Beem${}^{b}$}
\vskip.4cm
}

\centerline{\it Center for Theoretical Physics}
\centerline{\it University of California, Berkeley}
\centerline{\it Berkeley, CA 94720, USA}

\abstractmod{The disk partition function of the open topological string computes the spacetime superpotential for D-branes wrapping cycles of a {\it compact} Calabi-Yau threefold. We use string duality to show that when appropriately formulated, the problem admits a natural geometrization in terms of a {\it non-compact} Calabi-Yau fourfold without D-branes. The duality relates the D-brane superpotential to a flux superpotential on the fourfold.  This sheds light on several features of superpotential computations appearing in the literature, in particular on the observation that Calabi-Yau fourfold geometry enters the problem. In one of our examples, we show that the geometry of fourfolds also reproduces the D-brane superpotentials obtained from matrix factorization methods.}

\Date{\vbox{\hbox{\sl {\hskip0.0cm October 13, 2009}}
\vskip-.2cm
\hskip-.7cm{\bf ---------------------------------}
\vskip-.3cm
\noi {\ninerm \hbox{\hskip.035cm \tt ${}^a\;\,$mina@math.berkeley.edu}}
\vskip-.15cm
\noi {\ninerm \hbox{\hskip.035cm \tt ${}^b\;\,$beem@berkeley.edu}}
}}
\goodbreak
\vfill
\eject

\newsec{Introduction}

Much of the progress in our understanding of topological string theory on Calabi-Yau threefolds has been driven by its numerous intersections with physical superstring theory. For a non-compact Calabi-Yau, input from string dualities led to a computation of both open and closed topological string amplitudes to all orders in perturbation theory by means of the topological vertex \TV. Recently, these results have been verified by mathematicians \MNOP.
In a lesser measure, progress has also been made with regard to topological string amplitudes on {\it compact} Calabi-Yau manifolds. For example, closed topological string amplitudes have been computed in perturbation theory up a very high genus in \HuangHQ.

For the open topological string on the quintic, the disk amplitude has been computed for an involution brane in a ground-breaking paper by Walcher \Walcher.  In this case there are no massless open-string deformations, and the disk amplitude depends on closed-string moduli alone. The results of \Walcher\ have subsequently been verified by mathematicians in \PSW. More such examples were studied in \refs{\KW,\Knapp,\WalcherUJ}, the results of which were formalized in \MW\ within the framework of Griffiths' normal functions \Griffiths.

Despite the successes of \Walcher\ and subsequent papers, a more general framework is desirable. In particular, the topological string disk amplitude can depend on massless open-string moduli -- a situation which lies outside the scope of \refs{\Walcher,\MW}. Moreover the topological string on a disk computes the D-brane superpotential. The superpotential, corresponding to the classical brane action in topological string field theory, is naturally an off-shell quantity. For example, the superpotential for a B-brane wrapping a curve is given by
\eqn\first{
W({\cC}) = \int_{B({\cC})} \Omega^{(3,0)}
}
where ${\cC}$ is {\it any} curve, not necessarily holomorphic, and $B(\cC)$ is a three-chain with $\cC$ as its boundary. The critical points of \first\ with respect to variations of the brane embedding are holomorphic curves. Restricting to the critical locus, one recovers the normal functions of \refs{\Walcher,\MW}.

A method for computing the off-shell superpotential \first\ for ``toric branes'', first defined in \AV, has been proposed in \refs{\LercheCK,\LercheYW}, following \refs{\Mayrone,\LercheCW,\MayrZI}, and extended to compact Calabi-Yaus in \refs{\JS,\Mayr,\JockersMN}. For these branes, the superpotentials (as well as the open-string flat coordinates) are the solutions to a system of ``open/closed'' Picard-Fuchs equations, which arise as a consequence of ``$\cN=1$ special geometry''. For closed string, the Picard-Fuchs equations can be read off from the associated gauged linear sigma model (GLSM). The authors of \Mayr\ extend this formalism and associate an auxiliary GLSM to the open/closed Picard-Fuchs system, thus treating open and closed-string moduli at the same footing.

While these results are remarkable, their physical underpinnings remain somewhat mysterious. In formulating the Picard-Fuchs system, one must specify a divisor in the B-model geometry. This divisor is only a part of the combinatoric data which enters into the definition of the curve $\cC$ as a toric brane. The role of this divisor then requires some explanation. In addition, the appearance of the auxiliary GLSM begs for a physical interpretation. Finally, the methods of \Walcher\ extend beyond the class of toric D-branes, and so there should be some appropriate generalization of the techniques of \refs{\LercheCK,\LercheYW} which allows for the treatment of these other cases.

In this paper, we show that duality of the physical superstring explains these remarkable results. Consider the theory obtained by wrapping a D3 brane on a holomorphic two-cycle ${\cC}$ in the compact threefold $X_3$. The theory on the brane has ${\cal N}=(2,2)$ supersymmetry in two dimensions, with the superpotential \first\ computed by the disk partition function of the topological B-model with boundary on $\cC$. We will argue that the same superpotential is generated by a modified brane configuration, with an additional D5 brane wrapping a divisor $D$, and the D3 brane dissolved as world-volume flux. Note that this is {\it not} a duality; the modified configuration only produces the same answers for a certain subset of physical quantities. In particular, the moduli space for $D$ is not equivalent to the configuration space of the curve $\cC$. The superpotentials of the two theories must agree only for those variations of $\cC$ which are encoded in variations of the moduli of the divisor. In fact, given $X_3$ and ${\cC}$ we argue that the superpotential is the same for any configuration of D3, D5 and D7 branes where the D3 charge brane ends up localized on ${\cC}$.

The superpotential is also the same for any other D-brane configuration related by dimensional reduction/oxidation in ${\bf R}^{3,1}$. Fore example, the superpotentials for a D3 brane and a D5 brane wrapping the curve $\CC$ are the same. Our choice is such that the D-branes have codimension two in ${\bf R}^{3,1}$. For D-branes of lower codimension, only a subset of these models are consistent due to RR tadpoles. Tadpole cancellation in these cases requires the introduction of ingredients, such as orientifold planes, that are extraneous to the problem at hand.\foot{For example, D5 branes wrapping curves on a compact Calabi-Yau can be dissolved in the D7 branes, after introducing an appropriate number of orientifold planes so that the net D-brane charge vanishes, or by working in F-theory on a compact Calabi-Yau fourfold.} For higher codimension, the branes break more spacetime symmetries, complicating the problem. For this reason, the superstring embedding we have chosen is the most natural.

$S$-duality of type IIB string theory relates the D5 branes on $D$ to NS5 branes, with the flux remaining invariant since it is generated by a dissolved D3 brane. By further compactifying and $T$-dualizing one of the directions transverse to both $X_3$ and the NS5 branes, the branes are geometrized. The resulting configuration is then type IIA on a {\it non-compact} Calabi-Yau fourfold $X_4$. The flux on the NS5 brane is $T$-dualized to RR four-form flux $G_4$ on $X_4$, which generates a flux superpotential of the form,
\eqn\second{
W = \int_{X_4} G_4 \wedge \Omega^{(4,0)},
}
where $\Omega^{(4,0)}$ is the holomorphic four-form. Duality and the BPS nature of the superpotential guarantee that the superpotential \second\ is the same as \first\ as a function on appropriate moduli space. Moreover, since \second\ can be expressed in terms of closed-string periods of a Calabi-Yau fourfold, it is clear that the superpotential will satisfy a system of Picard-Fuchs equations which also encode the appropriate flat coordinates. This explains the appearance of the auxiliary toric data of Calabi-Yau fourfolds in \refs{\Mayrone, \Mayr}. Our approach is more general, however, allowing one to go beyond the category of toric branes.

The formalism of \refs{\LercheCK,\LercheYW,\JS,\Mayr,\JockersMN}, then, does {\it not} strictly reflect the physics of B-branes wrapping curves, but rather that of B-branes wrapping divisors $D$ with non-trivial world-volume flux.\foot{A different proposal for how to geometrize the D-brane superpotential by blowing up the divisor $D$, resulting in a threefold with is not Ricci-flat, has been proposed recently in \GrimmDQ.} In particular, to extract the disk amplitude for a B-brane on $\cC$, one must specify not only the divisor, but the first Chern class of the gauge bundle as well; on the Calabi-Yau fourfold, this corresponds to the choice of the RR flux. However, the distinction is for most purposes immaterial, as long as one is only interested in the superpotential.

The paper is organized as follows. In section two, we discuss the relation between D3 branes on $\cC$ and D5 branes on divisors containing $\cC$.  In particular, we show that upon the introduction of the appropriate fluxes on the D5 branes, the superpotentials for the two configurations agree. We then turn the chain of dualities that relates the D-brane geometry in IIB to a IIA flux compactification on a non-compact Calabi-Yau fourfold, which we give explicitly. We explain the role of mirror symmetry for Calabi-Yau threefolds and fourfolds in this context. In section three, we present detailed computations for a number of examples, which illustrate a variety of circumstances in which our prescription is of use. Among other things, we show that we can reproduce earlier results of \gaberdiel\ obtained from matrix factorizations. In an appendix, we discuss the relation of the methods developed here to the toric geometry approach of \Mayr.\foot{While this paper was in preparation for submission, two related works appeared \refs{\AlimBX,\GrimmEF}.
}

\newsec{B-brane Superpotentials on Calabi-Yau Threefolds}

Consider a D3 brane which wraps a curve $\cC$ inside a Calabi-Yau threefold $X_3$. This gives, in the non-compact directions, a $1+1$ dimensional theory with $\cN=(2,2)$ supersymmetry. When ${\cal C}$ is genus zero, there are no bundle moduli associated with the gauge fields on the D3 brane, and so any light degrees of freedom will arise from variations of the D-brane embedding. The number of massless chiral fields is equal to $H^0(\cC,\cN_{\cC})$, the number of holomorphic sections of the normal bundle to the curve in $X_3$, which encode infinitesimal, holomorphic deformations of the curve. On general grounds, the normal bundle splits as $\cN_C=\cO(-1+n)\oplus\cO(-1-n)$. Since an $\cO(k)$ bundle has $k+1$ holomorphic sections, the number of massless deformations for the curve is $n$.  For $n=1,$ there is one massless adjoint chiral field. This does not imply that the curve has finite holomorphic deformations -- there may be obstructions at higher order. Such an obstruction is encoded by a superpotential for the moduli \ShamitB. This superpotential is computed at string tree-level by the topological B-model, with boundaries on $\cC$. Alternatively, this amplitude can be computed in terms of the classical geometry of the brane configuration \AV,
\eqn\chain{
W(\cC)=\int_{B(\cC,\cC_*)}\;\Omega^{(3,0)},
}
where $\Omega^{(3,0)}$ is the holomorphic three-form on $X_3$, and $B(\cC,\cC_*)$ is a three-chain with one boundary on $\cC$, and the other on a homologous, reference two-cycle $\cC_*$.

In the generic case, the normal bundle to a curve is $O(-1)\oplus O(-1)$. This  corresponds to $n=0$, and there are no massless fields on the brane. The B-model disk amplitude then depends on the closed-string moduli alone, i.e., it measures the superpotential \chain\ evaluated at its critical point with respect to open-string variations. This scenario has been studied in \Walcher. The dependence of the physical D-brane superpotential on massive brane deformations can nevertheless be interesting and relevant to the low-energy effective theory above a certain scale \ShamitB.

In practice, evaluating these chain integrals directly from first principles is difficult. For the mirrors of non-compact, toric Calabi-Yau threefolds, the computations are rendered tractable by the relative simplicity of the geometry. This is not the case when the Calabi-Yau is compact, and so we will instead explore an alternate approach.

There are other brane configurations in string theory that give rise to the same superpotential \chain. For example, we may consider a D5 brane which wraps a divisor $D\subset X_3$. The divisor has $h^{2,0}(D)$ complex moduli. Each such modulus corresponds to a massless chiral field on the D5 brane. This moduli space is lifted when there is non-trivial flux on the D5 brane world-volume. In particular, take the flux $F$ to be \Poin\ dual to a curve $\cC\subset D,$
\eqn\pd{
F={\rm PD}_D[\,\cC\,].
}
For a generic D5 brane embedding, supersymmetry will be broken by the flux. A condition for unbroken supersymmetry is that the gauge bundle on the brane be holomorphic,\foot{There is an additional supersymmetry constraint on $J\wedge F$, where $J$ is the \Kah\ form on the four-cycle \MarinoAF.
This is interpreted in four dimensions as a D-term constraint, and so does not affect the superpotential.}
\eqn\flux{\eqalign{
F^{(0,2)}=0.
}}

This condition is equivalent to the requirement that the curve $\cC$ be holomorphic -- the same requirement that is enforced at the critical points of \chain. In fact, it can be shown that the flux \pd\ generates {\it precisely} the superpotential \chain\ for an appropriate subset of open-string deformations. The superpotential due to the flux \pd\ can be written \refs{\GomisWC,\MartucciIJ,\DenefVG,\DenefWQ,\CollinucciPF}
\eqn\Divsup{
W(D)=\int_{\Gamma(D,D_\star)}F\wedge\Omega^{(3,0)},
}
where $\Gamma$ is a five-chain which interpolates between $D$ and a homologous reference divisor $D_\star$ and $F$ is the appropriate extension of the flux as a closed form onto $\Gamma$ (obtained by taking the \Poin\ dual to $B(\cC,\cC_\star)$).  By \Poin\ duality, this superpotential can be rewritten as
\eqn\chaintwo{
W(D)=\int_{B(\cC,\cC_\star)}\Omega^{(3,0)},
}
which matches \chain\ for deformations which are common to the two brane systems.

Alternatively, one could consider a D7 brane which wraps all of $X_3$. The superpotential for the brane is the holomorphic Chern-Simons functional,
$$
W({X_3}) = \int_{X_3} A\wedge {\bar \del}A \wedge \Omega^{(3,0)}.
$$
Consider now turning on world-volume flux, $F\wedge F$, which is \Poin\ dual to the curve $\cC$,
\eqn\pdt{
F\wedge F={\rm PD}_{X_3}[\,\cC\,].
}
It is easy to see \AV\ that the D7 brane superpotential is the same as \chain . Namely, locally, near $\cC$ we can write
$$\Omega^{(3,0)} = d \omega
$$ 
for a two form $\omega$. Integrating by parts and using \pdt\ we can write 
$$
W(X_3) = \int_{\cC} \omega,
$$ 
which is the same, up to a constant as \chain.

From the superstring perspective, the superpotentials \chain, \Divsup, \pdt\ are all identical since they have the same have the same origin -- the D3 brane charge that is supported on $\cC$. For example, world-volume flux on a D5 brane of the type described above carries D3 brane charge due to a Wess-Zumino coupling on the brane world-volume of the form
\eqn\WZ{
S_{WZ} \qquad \sim \qquad\int_{\rm D5} F \wedge C^{(4)},
}
where $C^{(4)}$ is the four-form RR potential. When $F$ is as in \pd, this reduces to
\eqn\WZA{
S_{WZ} \qquad \sim \qquad \int_{\cC\times{\bf R}^{1,1}} C^{(4)},
}
so the D5 brane carries the charge of a D3 brane wrapping $\cC$. Similarly, turning on \pdt\ on the D7 brane, gives it a charge of a D3 brane supported on $\cC$.

\subsec{String duality and Calabi-Yau fourfolds}

The reformulation of the superpotential computation in terms of a D5 brane wrapping a divisor is of great use due to a duality which relates the problem to the classical geometry of Calabi-Yau fourfolds.

First, type IIB $S$-duality exchanges D5 branes and NS5 branes, leaving D3 branes invariant. So, we could have equivalently obtained \Divsup\ as the superpotential for an NS5 brane on the divisor $D$. Even though $S$-duality exchanges strong and weak coupling, the superpotential remains invariant when we compare the two theories at weak coupling. One way to see this is to note that the supersymmetry constraints \flux, which are reproduced by the superpotential \Divsup, are the same for both D5 and NS5 branes. Next, we compactify and $T$-dualize on one direction of ${\bf R}^{3,1}$ transverse to the NS5 brane. $T$-duality on this circle relates IIB to IIA and geometrizes the NS5 branes. The resulting geometry preserves $1+1$ dimensional Lorentz invariance and $\cN=(2,2)$ supersymmetry, so it is a Calabi-Yau fourfold, which we denote by $X_4$. Since one of the two directions transverse to the NS5 brane remained non-compact, the fourfold is non-compact as well.

The fourfold $X_4$ can be described explicitly as follows \refs{\D,\KMP,\AK}. In $X_3$, to the divisor $D$ is associated a line bundle $\cL_D$ over $X_3$ and a section $f_D$, such that $D$ is the zero-locus of the section,
\eqn\Div{
f_D(x)=0.
}
The Calabi-Yau $X_4$ is a $\;\IC^\star$ fibration over $X_3$ which degenerates over $D$, and can be described globally as
\eqn\fiber{
uv = f_D.
}
Here $u,v$ are sections of line bundles $\cL_1$ and $\cL_2$ over $X_3$, where $\cL_1 \otimes \cL_2 = \cL_D$. The locus where $u$ or $v$ go to infinity is deleted, so the manifold is non-compact. The fiber over a given point of $X_3$ is a copy of $\;\IC^\star$ described by
$$
uv = {\rm const.}
$$
This fiber has the topology of a cylinder, and is mirror to the ${\bf R}\times S^1$ formed by the two directions transverse to the NS5 brane and to $X_3$. It degenerates to $uv=0$ over the divisor $D$ that was wrapped by the NS5 brane.\foot{The correspondence between the open/closed-string geometry of the D3/NS5 brane system on $X_3$ and the geometry of $X_4$ is related to the duality of IIB on Calabi-Yau orientifolds to F-theory on Calabi-Yau fourfolds \DenefWQ. In the present case, $X_4$ is a fibration over $X_3$ with {\it non-compact} fiber. The main simplification here is that, due to the low codimension of branes, there are no tadpoles on $X_3$ to begin with. Correspondingly, the dual fourfold is non-compact.}

There are several immediate and important consequences of this correspondence. First, the moduli of the divisor entering into the choice of section $f_D$ become complex structure moduli of $X_4$. The holomorphic three-form on $X_3$ lifts to the holomorphic four-form on $X_4$,
$$
\Omega^{(4,0)} = \Omega^{3,0} \wedge du/u.
$$
The compact, Lagrangian four-cycles of $X_4$ come in two different flavors.  First, every closed Lagrangian three-cycle lifts to a Lagrangian four-cycle when combined with the $S^1$ in the fiber.  In addition, there are four-cycles which project to three-chains in $X_3$ with boundaries on $D$.  The generic fiber over the chain is still an $S^1$, but the circle now degenerates over $D$, capping off a closed cycle in $X_4$. Note that this means that the Lagrangian $T^4$ fibration of $X_4$ is related to the Lagrangian $T^3$ fibration of $X_3$ by simple inclusion of the $S^1$ fiber.

These observations imply that the superpotential \chain\ can be represented, on the Calabi-Yau fourfold $X_4$, as the period of $\Omega^{(4,0)}$ over an appropriately chosen four-cycle $L_4(B)$ which is the $S^1$ fibration over the three-chain $B(\cC,\cC_\star)$,
\eqn\foursuper{
W = \int_{L_4(B)} \Omega^{(4,0)}.
}
Such a superpotential for the complex structure moduli can only be generated by the presence of RR four-form flux $G_4$ on $X_4$. The flux generated superpotential is \GVW\
\eqn\HS{
W_{flux} = \int_{X_4} G_4 \wedge \Omega^{(4,0)}.
}
We now show that $T$-duality indeed implies that flux on the NS5 branes maps to the four-form flux of just the right value that $W_{flux}$ is equal to $W$ in \foursuper, thus providing a check for the duality.

One way to study the superpotential generated by fluxes is to study the corresponding BPS domain walls \GVW. To begin with, the superpotential \chain\ in IIB on $X_3$ is generated by world-volume flux $F$ on the D5 brane which is supported on a curve $\cC$. Different vacua correspond to different curves $\cC_i \subset D$ which are homologous in $X_3$, but distinct in $D$. If $B(\cC_1,\cC_2)$ is the three-chain that interpolates between two curves, the relevant domain wall is a D3 brane which wraps $B(\cC_1,\cC_2)$, with boundaries on the D5 brane. Under $S$/$T$-duality, the D3 brane domain wall becomes a D4 brane wrapping a special Lagrangian four-cycle $L_4(B)$ in $X_4$, obtained as the $S^1$ fibration over $B(\cC_1,\cC_2)$. This domain wall interpolates between vacua where RR four-form flux shifts by an amount \Poin\ dual to $L_4(B)$,
\eqn\G{
G_4 = {\rm PD}_{X_4}[L_4(B)].
}
Inserting this in $W_{flux}$ \HS\ we precisely recover \foursuper.

Thus the problem of computing the open-string superpotential \chain\ is rephrased as determining the periods of the holomorphic four-form on $X_4$ which control the flux superpotential \HS. Before proceeding to the calculation of such periods in explicit examples, however, we discuss the role played by mirror symmetry for the Calabi-Yau fourfolds in these geometries.

\subsec{$T$-duality and mirror symmetry for fourfolds}

Mirror symmetry provides another piece of evidence for the proposed correspondence. We recall in advance that mirror symmetry for Calabi-Yau $n$-folds can be interpreted as $T^n$ duality on Lagrangian $T^n$ fibers \SYZ.

To begin with, consider IIB on $X_3$ with a D3 brane wrapping $\cC$ {\it before} adding D5 or NS5 branes. Mirror symmetry for Calabi-Yau threefolds is a $T^3$-duality on the special Lagrangian $T^3$ fibers of $X^3$. This maps $X_3$ to its mirror ${Y}_3$, IIB to IIA, and D3 branes on $\cC$ to D4 branes wrapping a Lagrangian three-cycle $L$. Mirror symmetry for threefolds also exchanges the topological A- and B-models, so the superpotential for the low-energy effective theory on the D4 branes is computed by the disk amplitude of the topological A-model. This amplitude receives contributions from holomorphic maps of a worldsheet with the topology of a disk into $Y_3$, with boundaries on $L$. For $b_1(L)=n$, there are $n$ non-contractible one-cycles in $L$, which are contractible in $Y_3$ since $b_1(Y_3)=0$. The one-cycles can then be filled in to disks in $Y_3$. Let $u$ denote complexified \Kah\ volume of a minimal area disk, and $t$ the closed-string \Kah\ modulus. The at large radius in both closed and open-string moduli, the disk amplitude has the form
\eqn\disk{
W = P_2(u, t)+ \sum_{n=1}\sum_{q, Q} \, {N_{q,Q}\over n^2}\; e^{- n (q u + Qt)}
}
where $(q,Q)$ denotes the relative homology class of the disk, the sum over $n$ is a sum over multi-covers and $N_{q,Q}$ are {\it integers}. We have added a polynomial quadratic in $u$ to cover the case when the moduli are actually massive, i.e., the Lagrangian has the topology of an $S^3$ and $b_1(L)=0$. There can still be holomorphic disks ending on the brane, whose \Kah\ moduli $u$ are expressible in terms of the \Kah\ moduli $t$ of the Calabi-Yau, corresponding to the extrema of the quadratic part of the superpotential. An example of this in the non-compact setting was given in \AVG.

Now consider adding NS5 branes on the divisor $D$ to the IIB setup on $X_3$. If we perform a $T^3$-duality on the Lagrangian $T^3$ fibers now, the NS5 branes will again be geometrized. This is because a $T$-duality on an odd number of circles transverse to the NS5 branes geometrizes them. Here, the NS5 branes wrap a divisor, so only one of the $T$-dualized circles is transverse to them.  The theory preserves the $1+1$ dimensional Lorentz invariance and ${\cal N}=(2,2)$ supersymmetry, and so the dual geometry is again a Calabi-Yau fourfold. In fact, this fourfold $Y_4$ is nothing but the mirror of $X_4$! To see this, note that the $T^4$ fiber of $X_4$ is an $S^1$ fibration over the $T^3$ fiber of $X_3$. The statement then follows upon refining mirror symmetry on the fourfold as $T^4$-duality in two steps: a $T$-duality relating $X_4$ to NS5 branes on $X_3$, followed by a $T^3$-duality relating this to IIA on $Y_4$.

So, after introducing an NS5 brane on the divisor $D$, the mirror of $X_3$ is no longer the Calabi-Yau threefold with Lagrangian D4 brane, but rather is a Calabi-Yau fourfold with flux. However, we have argued in this note that the physics of the superpotential must remain the same. We'll now show how the superpotential \disk\ is reproduced in the mirror. Mirror symmetry relates complex structure moduli of $X_4$ to \Kah\ moduli of $Y_4$. A superpotential for these \Kah\ moduli can be generated by $0-$, $2-$, $4-$, $6-$ and $8-$form fluxes  \refs{\GVW,\GukovIQ,\MayrSH}. To determine the superpotential, we again follow the BPS domain walls through the chain of dualities. In the context of IIB on $X_3$ with NS5 branes on $D$, the domain walls were D3 branes wrapping special Lagrangian three-chains $B(\cC,\cC')$ and ending on the NS5 branes. The $T^3$-duality that maps to IIA on $Y_4$ sends the D3 branes to D4 branes on a four-chain $D_4(B)$ that interpolates between the Lagrangians $L_1$ and $L_2$ which are mirror to $\cC_1$ and $\cC_2$. The RR flux that shifts across these is a four-form flux, \Poin\ dual to $D_4(B)$,
\eqn\fluxb{
G_4 = {\rm PD}_{Y_4}[D_4(B)].
}
This implies that the superpotential on $Y_4$ is
\eqn\VS{
W_{flux} =  \int_{Y_4} G_4 \wedge k \wedge k,
}
where $k$ is the complexified \Kah\ form. In particular, inserting the jump in the flux over the D4 brane domain wall \fluxb, we precisely recover the BPS tension of D4 brane on $L_4(B)$.

It is a remarkable fact, and a check of the duality chain proposed here, that the flux superpotential \VS\ in the fourfold has, at large radius, the integral expansion \disk\ \refs{\GMP,\MayrSH}. For this, it is crucial that there are only four-form fluxes turned on, which is exactly what is needed for the theory to be dual to IIA on Calabi-Yau threefold $Y_3$ with D4 branes wrapping a Lagrangian three-cycle $L$.

Before we turn to examples, note that we have made no a-priori restriction on $D$. In particular, the divisor $D$ does not have to be a ``toric'' divisor. When $D$ is toric (and $X_3$ a hypersurface in a toric variety), it is easy to see that the Calabi-Yau fourfold \fiber\ has the same complex structure as those in \Mayr , so our results are guaranteed to agree.

\newsec{Examples}

In this section, the observations of the previous sections are applied to a number of brane configurations on families of Calabi-Yau threefolds. The first examples will focus on D3 branes wrapping degree-one rational curves on the quintic in the vicinity of the Fermat point. These brane configurations have been studied in \gaberdiel\ using matrix factorization/worldsheet CFT methods, and the results reported here will agree with those obtained previously. The second set of examples draws from the class of ``toric branes'' near the large complex structure point of the mirror quintic. These branes have previously been discussed in \refs{\Mayr}. The present treatment will agree with those results, but will provide a new interpretation for some of the methods involved in the calculations. Finally, we consider several examples of branes on complete intersection Calabi-Yaus which were studied recently in \WalcherUJ.

In all of these examples, our approach will be results-oriented and will not include an exhaustive analysis of the fourfold geometries involved.  In particular, we will not attempt to derive the complete set of Picard-Fuchs operators for the fourfolds, and will instead settle for a set of differential operators which uniquely determines the periods of interest given the desired leading-order behavior. We will borrow the overall normalization of the superpotentials from results elsewhere in the literature, leaving a careful intersection/monodromy analysis for future study.

\subsec{D-branes on the Fermat quintic}

The starting point for this first set of examples is the Fermat quintic, given by the hypersurface in $\IP^4$,
$$
\fermat55555=0.
$$
There are continuous families of rational curves on this quintic, which we specify by their parameterizations in terms of homogeneous coordinates $(u,v)$ on a $\IP^1$ as
\eqn\family{\eqalign{
(\x1,\x2,\x3,\x4,\x5)&=(au,bu,cu,v,-\eta v)\cr
a^5+b^5&+c^5=0,
}}
where $\eta$ is a fifth root of unity. There are fifty such families, corresponding to ten partitions of the $x_i$ into groups of two and three, and five values of $\eta$. For later convenience, we denote these families thusly,
\eqn\fams{
\Sigma^{ijk}_{m}(a,b,c):\quad (x_i,x_j,x_k)=(au,bu,cu),\qquad\eta=e^{2\pi im/5}.
}
The coefficients $(a,b,c)$ are only defined up to an overall scaling, so each family is parameterized by a complex one-dimensional curve which is a hypersurface in $\IP^2$. Each family of curves intersects another family at points where one of the homogeneous coordinates vanishes.

The presence of continuous families of rational curves is non-generic, and perturbing the bulk complex structure will lift (some of) the families, leaving only isolated curves. Such a perturbation should therefore generate a superpotential for D3 branes which wrap these cycles. Such a scenario was explored in \gaberdiel, where worldsheet CFT techniques were used to compute the superpotential at leading order in such a bulk perturbation as an analytic function on the open-string moduli space.

In order to compute the superpotential for one of these families of curves using the methods of section two, a family of divisors must be chosen such that each divisor is transverse to the family of curves, and each member of the family of curves is subsumed by a single member of the family of divisors. In particular, choosing divisors
\eqn\div{
D(\phi):\qquad x_4+\phi x_5=0,
}
the Picard-Fuchs equations for the associated fourfold will encode the superpotential for any of the fifteen families $\Sigma^{i45}_m$. For a given value of $\phi$, this divisor encompasses the curve with $\phi=-b/c$, and so allows for a good parameterization of the family away from the points at $c=0$.\foot{By making the change of variables $\phi\rightarrow\phi\inv$, one can equally well recover the physics in the neighborhood of $c=0$.}

Following \gaberdiel, we further consider perturbations to the complex structure of the threefold which are of the form
\eqn\quintperturbed{
\fermat55555+x_1^3g(x_3,x_4,x_5)
}
where
\eqn\perturbation{
g(x_3,x_4,x_5)=\sum_{p+q+r=2}g_{pqr}x_3^px_4^qx_5^r.
}
For simplicity, we restrict to monomial perturbations -- the case for more general perturbations can be treated by the same methods. In particular, the following two bulk perturbations will lead to qualitatively different physics on the D3 branes,
\eqn\twoex{
g_1(x_3,x_4,x_5)=\psi_1\,x_4x_5\qquad\qquad g_2(x_3,x_4,x_5)=\psi_2\,x_4^2.
}

The superpotential for the family of D3 branes is encoded in the periods of a non-compact Calabi-Yau fourfold $X_4$. The line bundles introduced in section two to facilitate construction of the fourfold construction can be identified simply as the bundles $\cO(n)\rightarrow\IP^4$ restricted to $X_3$,
\eqn\bundles{\eqalign{
\cL_D&=\cO(1)\rightarrow\IP^5|_{X_3},\cr
\cL_1&=\cO(1)\rightarrow\IP^5|_{X_3},\cr
\cL_2&=\cO(0)\rightarrow\IP^5|_{X_3}.
}}
It follows that $X_4$ is a complete intersection in $\IP^6\times\;\IC$ given by
\eqn\fourfold{
Q(\psi_i)=0\qquad\qquad P(\phi)=x_6x_7+D(\phi)=0,
}
with the points $(0:0:0:0:0:x_6\,;\,x_7)$ deleted. The periods of $\Omega^{(4,0)}$ for these geometries can be computed using standard methods, as we now summarize.

For the first bulk perturbation $g_1(x_3,x_4,x_5)$, the D3 brane moduli spaces for the families $\Sigma^{245}_m,\;\Sigma^{345}_m$ are lifted, with the only remaining holomorphic curves in these families being located at the points in \family\ where
\eqn\firstcrit{
b\cdot c=0.
}
This leads to five distinct curves at $b=0$ and another five at $c=0$. The divisors \div\ provide a description of the configurations with $b=0$. The normal bundle to these curves is $\cN_\Sigma=\cO(-1)\oplus\cO(-1)$, rendering the theory on the D3 branes massive.

Picard-Fuchs operators\foot{In all examples, we use the scaling symmetries of the ambient projective space as an efficient way to produce GKZ-type operators which are guaranteed to annihilate the compact periods of the fourfold.} for the fourfold \fourfold\ can be derived using the residue representation for periods of the holomorphic four-form \Griffiths,
\eqn\ResPer{
\Pi_\alpha(\psi,\phi)=\int_{\gamma_1\times\gamma_2\times\Gamma_\alpha}{\Delta\over Q(\psi_1)P(\phi)},
}
where $\gamma_1\times\gamma_2\times\Gamma_\alpha$ is a tubular neighborhood constructed about the desired four-cycle,
$$
\Delta=\sum_{i=1}^6(-1)^iw_ix_idx_1\wedge\ldots\wedge\widehat{dx_i}\wedge\ldots\wedge dx_6\wedge dx_7,
$$
and the $w_i$ are the scaling dimensions of the homogeneous coordinates. We initially represent the hypersurface in terms of redundant parameters on the fourfold moduli space,
\eqn\redun{
Q(a_i)=\sum_{i=1}^5a_ix_i^5+a_0\x1^3\x4\x5\qquad P(b_i)=x_6x_7+b_4x_4+b_5x_5.
}
where the algebraic coordinates on the moduli space can be given in terms of these parameters by a rescaling of the $x_i$,
$$
\psi_1={a_0\over a_1^{3/5}a_4^{1/5} a_5^{1/5}},\qquad\qquad \phi={b_5a_4^{1/5}\over b_4a_5^{1/5}}.
$$
The periods expressed in terms of the redundant parameters,
\eqn\redperiod{
\hat{\Pi}_\alpha(a_i,b_i)=\int_{\gamma_1\times\gamma_2\times\Gamma_\alpha}{\omega\over Q(a_i)P(b_i)},
}
can be related to the physical periods \ResPer\ according to
\eqn\perprelat{
\hat{\Pi}_\alpha(a_i,b_i)={1\over(a_1a_2a_3a_4a_5)^{1/5}}\Pi_\alpha(\psi_1,\phi).
}

It is easy to produce differential operators which annihilate the periods $\hat{\Pi}_\alpha(a_i,b_i)$ because differentiation with respect to the $a_i,b_i$ can be performed under the integral. As such, we obtain the following operators,
\eqn\diffopone{\eqalign{
L_1&=\del_{a_1}^3\del_{a_3}\del_{a_4}-\del_{a_0}^5,\cr
L_2&=\del_{b_3}^5\del_{a_5}-\del_{b_5}^5\del_{a_3}.
}}
These operators in turn are equivalent to relations on the periods written in terms of the algebraic moduli,
$$
\cL_i\Pi_{\alpha}(\psi,\phi_i)=0,
$$
where, after factorizing, the operators can be written as
\eqn\diffoptwo{\eqalign{
\cL_1&=\prod_{k=0,\hat{3}}^4(\th_{\psi_1}-k)+3\left({\psi_1\over5}
\right)^5(3\th_\psi+11)(3\th_\psi+1)(\th_\psi+\th_\phi+1)(\th_\psi-\th_\phi+1)\cr
\cL_2&=\th_\phi(\th_\phi-\th_{\psi_1}-1)-\phi^5\th_\phi(\th_\phi+\th_\psi+1).
}}
with $\theta_z\equiv z\del_z$.

Among the solutions to these Picard-Fuchs equations are four that depend only on $\psi$, and which correspond to solutions which would have arisen in a similar analysis of the periods of $X_3$. In addition, there are four $\phi$-dependent solutions, from amongst which we identify two of interest,
\eqn\periodsA{\eqalign{
t&=\ts\phi
\left(1+{1\over15}\phi^5+{7\over275}\phi^{10}+{77\over187500}\phi^5\psi_1^5+\ldots\right),\cr
\Pi_1&=\ts\phi^2\psi_1\left(1+{8\over35}\phi^5+{3\over25}\phi^{10}+{36\over15625}\phi^5\psi_1^5+\ldots\right).
}}
The first of these defines the flat open-string coordinate. The second, by its leading behavior, can be identified as the superpotential induced by D3 brane flux in the class of $\Sigma^{345}_0$.

These results are in agreement with those of \gaberdiel. In particular, the term in $\Pi_1$ which is linear in $\psi_1$ can be written in terms of a hypergeometric function as
\eqn\linear{
\Pi_1=(\psi_1\phi^2)\,{}_2F_1\ts\left({2\over5},{4\over5};{7\over5};\phi^5\right)+\ldots.
}
Up to the overall normalization, this exactly matches equation (3.14) of \gaberdiel\ for this choice of bulk deformation. It is clear from this derivation, however, that the physical basis for writing the superpotential is {\it not} in terms of $\phi$, but rather in terms of the flat coordinate $t$ which represents an appropriate period of $X_4$. Moreover, the overall normalization of the superpotential should be adjusted by the fundamental period of the fourfold as in conventional mirror symmetry calculations. We note that the fundamental period, $\Pi_0(z)$ for these fourfolds is {\it identical} to that for the related threefold -- in particular, it is independent of $\phi$ -- and so the change in normalization doesn't effect the superpotential at leading order in the $\psi_1$. However, for higher order corrections, it should be the normalized result which would match any CFT computation such as those performed in \gaberdiel. In light of these considerations, we display the {\it physical} superpotential,
\eqn\superphysone{
\cW(z,t)={\Pi_1(z,t)\over\Pi_0(z)}=t^2z\(\ts 1+{2\over21}t^5+{17\over31250}z^{5}+{2\over99}t^{10}+{68\over46875}t^5z^5+{38299\over128906250000}z^{10}+\ldots\)
}

The second bulk perturbation also lifts the D3 brane moduli space for the families $\Sigma^{245}_m$ and $\Sigma^{345}_m$ leaving only the holomorphic curves given by \family\ along with
\eqn\crittwo{
b^2=0.
}
For each family, this leads to five solutions, each of degeneracy two, at $b=0$. Thus, D3 branes wrapping these curves find themselves at the critical point of a higher-order superpotential. The normal bundle to each of these curves is $\cN_\Sigma=\cO(1)\oplus\cO(-3)$. However, only {\it one} of the holomorphic sections of the normal bundle is encoded by the variation of $\phi$ for the divisor.\foot{An analysis of the full moduli space of the divisor $D$ would encode all holomorphic deformations of the curves.} As a result, the following derivation produces the superpotential with respect to only one of the massless deformations.

As before, the differential relations for $\hat{\Pi}_{\alpha}(a_i,b_i)$ can be determined,
\eqn\relate{
\eqalign{
L_1&=\del_{a_1}^3\del_{a_4}^2-\del_{a_0}^5\cr
L_2&=\del_{a_3}\del_{b_5}^5-\del_{a_5}\del_{b_4}^3,
}}
where by rescaling the homogeneous coordinates the algebraic moduli for the fourfold can be found in terms of the redundant parameters,
$$
\psi_2={a_0\over a_1^{3/5}a_4^{2/5}},\qquad\qquad\phi={b_5a_4^{1/5}\over b_4a_5^{1/5}}.
$$
The resulting Picard-Fuchs operators which annihilate the periods $\Pi_{\alpha}(\psi_2,\phi_2)$ are (after factorizing) given by
\eqn\PFone{\eqalign{
\cL_1&=\prod_{k=0,\hat{3}}^4\left(\th_{\psi_2}-k\right)+3\left({\psi_2\over5}\right)^5(3\th_{\psi_2}+11)(3\th_{\psi_2}+1),(2\th_{\psi_2}-\th_\phi+6)(2\th_{\psi_2}-\th_\phi+1)\cr
\cL_2&=\th_\phi(\th_{\phi}-2\th_{\psi_2}-1)-{\phi}^5\th_\phi(\th_{\phi}+1).
}}
There are four $\phi$-{\it independent} solutions which are determined purely by the geometry of the threefold, and four additional $\phi$-dependent periods. Of these, the relevant flat coordinate and superpotential are given by
\eqn\pertwo{\eqalign{
t&=\ts\phi\left(1+{1\over15}\phi^5+{7\over275}\phi^{10}+\ldots\right),\cr
\Pi_2&=\ts\phi^3\psi_2\left(1+{3\over10}\phi^5+{54\over325}\phi^{10}+\ldots\right).
}}
It is worth noting that there is no solution which could correspond to a {\it massive} superpotential for $\phi$. The superpotential $\Pi_2$ can be written at leading order in $\psi_2$ in hypergeometric form,
\eqn\lineartwo{
\Pi_2=\ts(\psi_2\,\phi^3)\,{}_2F_1\left({3\over5},{4\over5};{8\over5};\phi^5\right)+\ldots
}
which agrees with equation (3.14) of \gaberdiel. The superpotential should again be expressed in the physical normalization in terms of flat open and closed-string coordinates, giving
\eqn\superphystwo{
\cW(z,t)={\Pi_2(t,z)\over\Pi_0(z)}=\ts t^3z\(1+{1\over10}t^5+{69\over31250}z^5+{248\over10725}t^{10}-{24\over78125}t^5z^5+{98999\over5371093750}z^{10}+\ldots\)
}

\subsec{Toric branes on the mirror quintic}

A class of D-branes which has been well studied in the context of open-string mirror symmetry are ``toric branes,'' first described in \AV, to which we defer for their description in terms of toric geometry. They were originally introduced as non-compact brane geometries in local Calabi-Yau threefolds, but more recently, progress has been made in understanding the extension to the compact case \refs{\JS,\Mayr}. We consider one of these examples from the perspective of the dual fourfold, and reproduce the same Picard-Fuchs equations and superpotentials which were derived in \Mayr\ based on a somewhat formal application of toric geometry/GLSM techniques.

The bulk geometry is the one-parameter family of quintic hypersurfaces in $\IP^4$ given by\foot{For our purposes, this can be either a one-dimensional slice of the complex structure moduli space of the quintic, or the $\Z_5^3$ orbifold of these geometries which constitutes the mirror quintic.}
$$
x_1^5+x_2^5+x_3^5+x_4^5+x_5^5+\,\psi\, x_1x_2x_3x_4x_5=0.
$$
The D3 brane geometries of interest are degree-two rational curves, with the following parameterization,
$$
(x_1,x_2,x_3,x_4,x_5)=(u^2,\a u^2,v^2,\b v^2,(-\a\b\psi)^{1/4}uv),
$$
where $\alpha^5=\beta^5=-1$. These are rigid curves, so there are no massless open-string moduli.

In \Mayr, these branes were described in the framework of toric geometry by identifying them as components of the intersection in the mirror quintic of two of the following three toric divisors,
$$
x_1^5+x_2^5=0,\qquad\quad x_3^5+x_4^5=0,\quad\qquad x_5^5+\psi\x1\x2\x3\x4\x5=0.
$$
By choosing one of these as the physical divisor for our prescription, we expect to reproduce the D3 brane superpotential with respect to certain massive open-string deformations as well as the bulk modulus $\psi$. As opposed to the previous examples, there are no privileged massive deformations that are singled out by the brane geometry. Consequently, there is no preferred choice for the physical divisor, and the off-shell superpotentials for the different branes will not match. However, the {\it on-shell} value of the superpotential with respect to the open-string degrees of freedom should be independent of the choice of divisor. Since the first and second divisors are related by a permutation symmetry, we will consider
\eqn\quintdiv{
D_1(\phi_1)=x_1^5+\phi_1x_2^5,\qquad\qquad D_2(\phi_2)=x_5^5+\phi_2x_1x_2x_3x_4x_5,
}
which contain information about the supersymmetric D3 branes at $\phi_1=1$ and $\phi_2=\psi$, respectively. The relevant line bundles for this configuration are\foot{Alternatively, one could choose $\cL_u=\cO(a)$ and $\cL_v=\cO(b)$ for any $a+b=5$. Such a choice does not affect the present discussion.}
\eqn\bundles{\eqalign{
\cL&=\cO(5)\rightarrow\IP^4|_{X_3},\cr
\cL_1&=\cO(5)\rightarrow\IP^4|_{X_3},\cr
\cL_2&=\cO(0)\rightarrow\IP^4|_{X_3}.
}}
The dual fourfold is a complete intersection in the weighted projective space $\IP^5_{111151}\times\;\IC$, with the points $(0:0:0:0:0:x_6,;\,x_7)$ removed. The defining equations are
\eqn\toricfour{
Q(\psi)=\sum_{i=1}^5x_i^5+\psi x_1x_2x_3x_4x_5=0,\qquad\qquad P(\phi_i)=x_6x_7+D_i(\phi_i)=0.
}
We now turn to the derivation and solution of the Picard-Fuchs equations for these fourfolds.

For the first divisor, $D_1(\phi_1)$, we proceeding in a manner analogous to the previous examples and find, after factorization, the following Picard-Fuchs operators,\foot{These can be obtained most easily as the GKZ-operators associated with the charge vectors which define the toric bulk/brane geometry, as in \Mayr. However, it is easy to see that the fourfold introduced above gives rise to the same operators.}
\eqn\torpf{\eqalign{
\cL_1&=\th_z^4-\th_z^2\th_{\phi_1}^2+z\prod_{k=1}^4(5\th_z+k),\cr
\cL_2&=\th_{\phi_1}(\th_{\phi_1}+\th_z)-\phi_1\,\th_{\phi_1}(\th_{\phi_1}-\th_z).
}}
where we expand about the large complex structure point of the mirror quintic, so have introduced $z=\psi^{-5}$. The holomorphic curves in question are located at $\phi_1=1$, and so we look for solutions to these equations expanded about that point, as a function of $\hphi=\phi_1-1$. There are precisely two solutions which are functions of $\hat{\phi}_1$ and finite at $z=0$,
\eqn\toricperP{\eqalign{
t&=\hphi-\half{\hphi^2}+\third\hphi^3(1-60z)-\quart\hphi^4(1-120z)+\ldots\cr
\Pi_1&=\ts\sqrt{z}\Bigl(1+{5005\over9}z+{52055003\over75}z^2+{283649836041\over245}z^3+{8908737478232449\over3969}z^4\cr
&\ts\hskip6.4cm\;\;+{1\over8}\hphi^2-{1\over8}\hphi^3+{15\over128}\hphi^4+\ldots\Bigr).
}}
In addition, there are four solutions which comprise the usual set of closed-string periods on the mirror quintic. After appropriately normalizing, these are the open-string flat coordinate and superpotential, respectively. This leads to the following expressions for the physically normalized superpotential as a function of the flat open/closed-string coordinates,
\eqn\toricsuperA{
\cW_{phys}(q,t)=\cW_{closed}(q)+{\ts{15\over8}q^{1/2}t^2\left(1-265q+\ldots\right)
}}
Where $\cW_{closed}$ is the {\it on-shell} superpotential as a function of the closed-string moduli when the open-string coordinate is fixed at its critical point,
\eqn\onshellw{
\cW_{closed}(q)=\ts15q^{1/2}+{2300\over3}q^{3/2}+{2720628\over5}q^{5/2}+{23911921125\over49}q^{7/2}+\ldots
}
This result precisely matches those of \refs{\Walcher,\Mayr}, which were obtained through a variety of different methods.

We now turn to the second divisor, $D_2(\phi_2)$.  This is precisely the divisor which was studied in \Mayr, and we reproduce the results here for completeness. Following \Mayr, we choose to work in terms of algebraic variables
$$
z_1=-\phi_2^{-1}\psi^{-4},\qquad\qquad z_2=-\phi_2\psi^{-1},
$$
with respect to which the Picard-Fuchs operators are given by
\eqn\picfuch{\eqalign{
\cL_1&=\th_1^5-\th_1^4\th_2-z_1\prod_{k=1}^4(4\th_1+\th_2+k)(\th_1-\th_2),\cr
\cL_2&=\th_1^2-\th_1\th_2-z_2(4\th_1+\th_2+1)(\th_1-\th_2),\cr
\cL_1^\prime&=\th_2\th_1^4+z_1z_2\prod_{k=1}^5(4\th_1+\th_2+k).
}}
The expected critical point is at $z_2=-1$, $z_1=\psi^{-5}$. To find a good expansion for the solutions to the Picard-Fuchs equations, we introduce coordinates
$$
u=z_1^{-1/4}(1+z_2),\qquad\qquad v=z_1^{-1/4},
$$
as a function of which the superpotential can be found in a power-series expansion,
\eqn\uvper{
\Pi_2={\ts{1\over8}u^2+15v^2+{5\over48}u^3v-{15\over2}uv^3+{1\over46080}u^6+{35\over384}u^4v^2-{15\over8}u^2v^4+{25025\over3}v^6}+\ldots
}
It can be verified that this superpotential has a critical point with respect to the open-string variation at $u=0$, as predicted by the geometry. Moreover, by setting the open-string deformation to zero, normalizing, and expressing the result in terms of the flat closed-string coordinate, the derived {\it on-shell} superpotential matches the results of \Walcher,
\eqn\walchsup{
\cW={\ts15q^{1/2}+{2300\over3}q^{3/2}+{2720628\over5}q^{5/2}+{23911921125\over49}q^{7/2}\ldots}
}
This matches the results obtained above using the first divisor \onshellw, although the physical theories on the different NS5 branes are inequivalent.

\subsec{$\IP^5[3,3]$}

Our next example is a one-parameter complete intersection Calabi-Yau, studied recently in \WalcherUJ. The geometry is mirror to the intersection of two cubics in $\IP^5$, and can be described as the quotient\foot{As in the examples on the mirror quintic, one can equally well consider this to be a special case of the larger A-mode geometry, $\IP^5[3,3]$.}
$$
X_3=\{W_1=0, W_2=0\}/G
$$
Where the $W_1$ and $W_2$ are the most general cubic polynomials invariant under the appropriate discrete symmetry group $G={\bf Z}_3^2\times{\bf Z}_9$,
$$\eqalign{
W_1&={x_1^3\over3}+{x_2^3\over3}+{x_3^3\over3}-\psi x_4x_5x_5\cr
W_2&={x_4^3\over3}+{x_5^3\over3}+{x_6^3\over3}-\psi x_1x_2x_3
}$$
The curves studied in \WalcherUJ\ are determined by the intersection of two hyperplane divisors in $X_3$,
$$
D_1=\{x_1+x_2=0\},\qquad\qquad D_2=\{x_4+x_5=0\}.
$$
This intersection is reducible, being comprised of one line and two degree four curves, the rational parameterizations of which can be found in \WalcherUJ.

We introduce a fivebrane on the divisor $D_1$, which we embed in the one-parameter family of divisors,
\eqn\CIdiv{
D_1(\phi)=x_1+\phi x_2
}
The line bundles for this configuration are then
\eqn\bundlesthree{\eqalign{
\cL_D&=\cO(1)\rightarrow\IP^5|_{X_3},\cr
\cL_1&=\cO(1)\rightarrow\IP^5|_{X_3},\cr
\cL_2&=\cO(0)\rightarrow\IP^5|_{X_3},
}}
and the dual fourfold is a complete intersection
\eqn\CIEqns{
W_1(\psi)=0,\qquad W_2(\psi)=0,\qquad P(\phi)=x_6x_7+D_1(\phi)=0,
}
in $\IP^6\times\;\IC$ with the points $(0:0:0:0:0:0:x_7\,;\,x_8)$ removed. Picard-Fuchs operators can be derived in the usual way, leading to
\eqn\CIPF{\eqalign{
\cL_1&=(\th_z^4-\th_w^2)\th_z^2-9z(3\th_z+2)^2(3\th_z+1)^2,\cr
\cL_2&=\th_w(\th_z+\th_w)+w\th_w(\th_z-\th_w),
}}
where we've introduced local variables $z=(3\psi)^{-6},\;w=\phi^3$. The holomorphic curves in question are located at $\phi=1$, so we find solutions expanded about the point $\hat{w}=w-1=0$. There are two $\hat{w}$-dependent solutions which are finite at $z=0$,
\eqn\CICYper{\eqalign{
t&=\hat{w}-\half \hat{w}^2+\third \hat{w}^3-\quart \hat{w}^4-6 z \hat{w}^3+\fifth \hat{w}^5+9 z \hat{w}^4+\ldots\cr
\Pi&=\Pi_{closed}(z)+{\ts\sqrt{z}\hat{w}^2({1\over8}-{1\over8}\hat{w}+{15\over128}\hat{w}^2-{1225\over384}z \hat{w}^2-{7\over64}\hat{w}^3+\ldots)}
}}
Where $\Pi_{closed}$ is the part of the superpotential which depends only on the closed-string moduli, i.e., the {\it on-shell} part that is accessible to the methods of \WalcherUJ.
\eqn\PIClosed{
\Pi_{closed}=\sqrt{z}{\ts\left(1+{1225\over9}z+{1002001\over25}z^2+{19200813489\over1225}z^3+{28214528710225\over3969}z^4\right).
}}
These periods are the flat open-string coordinate and superpotential, respectively. Again, the periods should be normalized by the fundamental period of $X_3$ and expressed in terms of the flat coordinates. The resulting superpotential is
\eqn\CISuper{
\cW(q,t)=\cW_{closed}+q^{1/2}t^2\left({9\over4}-{243\over2}q-{10935\over4}q^2+{3\over64}t^2+\ldots\right)
}
where, $\cW_{closed}$ is the {\it on-shell} superpotential obtained by setting $t\rightarrow0$,
\eqn\CIOnshell{
\cW_{closed}(q)={\ts 18q^{1/2}+182q^{3/2}+{787968\over25}q^{5/2}+{323202744\over49}q^{7/2}+{15141625184\over9}q^{7/2}}+\ldots
}
This matches equation (2.37) of \WalcherUJ.

\subsec{$\IP^5_{112112}[4,4]$}

As our final example, we consider another one-parameter complete intersection Calabi-Yau from \WalcherUJ.  This time, the A-model geometry is the intersection of two degree-four hypersurfaces in the weighted projective space $\IP^5_{112112}$.  The mirror geometry is given by the quotient
$$
X_3=\{W_1=0, W_2=0\}/G
$$
Where the $W_1$ and $W_2$ are the most general degree-four polynomials invariant under the appropriate discrete symmetry group $G={\bf Z}_2^2\times{\bf Z}_{16}$,
$$\eqalign{
W_1&={x_1^4\over4}+{x_2^4\over4}+{x_3^2\over2}-\psi x_4x_5x_5,\cr
W_2&={x_4^4\over4}+{x_5^4\over4}+{x_6^2\over2}-x_1x_2x_3.
}
$$
The curves in question are contained in the intersection in $X_3$ of the two divisors
$$
D_1=\{x_1^2+\alpha_1\sqrt{2}x_3=0\},\qquad\qquad D_2=\{x_2^2+\alpha_2\sqrt{2}x_6=0\}.
$$
where $\alpha_i=\pm i$.  The intersection consists of one line and three degree-five curves.

We introduce a fivebrane on the divisor $D_1$, which is embedded into the one-parameter family of divisors given by
\eqn\CIdiv{
D(\phi)=x_1^2+\phi x_3.
}
The line bundles for this configuration can be chosen as
\eqn\bundlesthree{\eqalign{
\cL_D&=\cO(2)\rightarrow\IP^5|_{X_3},\cr
\cL_1&=\cO(1)\rightarrow\IP^5|_{X_3},\cr
\cL_2&=\cO(1)\rightarrow\IP^5|_{X_3},
}}
and the dual fourfold is a complete intersection
\eqn\CIEqns{
W_1(\psi)=0,\qquad W_2(\psi)=0,\qquad P(\phi)=x_6x_7+D(\phi)=0,
}
in $\IP^8_{11211211}$ with the points $(0:0:0:0:0:0:x_7:x_8)$ removed. Picard-Fuchs operators can be derived in the usual way, leading to
\eqn\CIPF{\eqalign{
\cL_1&=(\th_z+\th_w)(2\th_z-\th_w-1)(2\th_z-\th_w)\th_z^2-16z(4\th_z+3)^2(4\th_z+2)(4\th_z+1)^2,\cr
\cL_2&=(\th_z+\th_w)\th_w-{1\over2}w(2\th_z-\th_w)\th_w,
}}
where we've introduced local variables $z=(8\psi)^{-4},\;w=\phi^2$. The holomorphic curves in question are located at $\phi=\pm i\sqrt{2}$, so we look for solutions expanded about the point $\hat{w}=w+1=0$.  Amongst the $\hat{w}$-dependent solutions, we identify the periods which correspond to open-string superpotentials,
\eqn\CICYpertwo{\eqalign{
\Pi_1&=\ts z^{1/3}(1+{8281\over16}z+{38130625\over49}z^2+{80263989481\over49}z^3+{1\over36}\hat{w}^2+{1\over81}\hat{w}^3+\ldots),\cr
\Pi_2&=\ts z^{2/3}(1+{559504\over625}z+{15557323441\over10000}z^2+{10590422929849\over3025}z^3+{1\over162}\hat{w}^3+\ldots).
}}
Along with the closed-string periods, we can compute the on-shell superpotential in terms of closed-string flat coordinates, finding
$$
\cW(q)=\ts24q^{1/3}+150q^{2/3}+{2571\over2}q^{4/3}+{417024\over25}q^{5/3}+{45420672\over49}q^{7/3}+{131074059\over8}q^{8/3}+\ldots
$$
Again, we've chosen the linear combination of solutions to match the results of \WalcherUJ.

\vskip.5cm
\centerline{\bf Acknowledgments}
\noindent M.A. is indebted to Cumrun Vafa for valuable discussions in 2001, shortly after \Mayrone\ appeared. The authors are grateful to Masoud Soroush for delivering a wonderful seminar at the BCTP which led to this line of research, as well as for subsequent discussion.  We would also like to thank Yu Nakayama for valuable discussions. This research was supported in part by the UC Berkeley Center for Theoretical Physics and NSF grant PHY-0457317.

\goodbreak
\vfill
\eject

\appendix{A}{Toric Methods}

In this appendix, we demonstrate the equivalence of the fourfolds derived by T-duality with those obtained using toric geometry/GLSM techniques in the case of toric branes \Mayr. We will take one of the branes of section 3.2 as our example. From the toric data which defines the brane geometry in question, the methods of \HV\ allow for a derivation of the appropriate B-model geometry.

Recall  that in the context of toric geometry, the quintic is described by the charge vector of the associated GLSM,
$$
Q_1=(-5, 1, 1, 1, 1, 1).
$$
Moreover, a certain class of ``toric branes'' can also be encoded in a similar charge vector \AV. In particular, for the charge vector
$$
Q_2=(1, -1, 0, 0, 0, 0),
$$
the mirror B-brane on the mirror quintic wraps the divisor
$$
x_1^5+\phi x_1x_2x_3x_4x_5=0.
$$
The approach of \refs{\Mayr,\LercheYW,\LercheCW} was to ``enhance'' these charge vectors to define an auxiliary GLSM for the open/closed-string geometry as follows
\eqn\vecs{\eqalign{
{Q}_1^\prime&=(-5, 1, 1, 1, 1, 1; 0, 0),\cr
{Q}_2^\prime&=(1, -1, 0, 0, 0, 0; 1, -1).
}}
This toric data then defines a system of GKZ differential operators. Defining coordinates on the complex structure moduli space for the B-model geometry according to
\eqn\toricmod{
z_a=(-)^{Q^\prime_{a,0}}\prod_{i=0}^7a_i^{Q^\prime_{a,i}},
} 
the differential operators which annihilate the periods of the holomorphic four-form are
\eqn\GKZ{
\cL_a=\prod_{k=1}^{\ell_0^a}(\th_{a_0}-k)\prod_{\ell_i^a>0}\prod_{k=0}^{\ell^a-1}(\th_{a_i}-k)-(-1)^{\ell_0^a}z_a\prod_{k=1}^{-\ell_0^a}(\th_{a_0}-k)\prod_{\ell_i^a<0}\prod_{k=0}^{-\ell_i^a-1}(\th_{a_i}-k),
}
where $\th_{a_i}=\sum_{a}Q^\prime_{a,i}\th_{z_a}$.  It is straightforward to check that the operators defined in this way match those obtained by our methods in section three.

Moreover, by applying the methods of \HV, we can further derive the B-model geometry which is defined by this toric data. Starting with the non-compact toric variety given by \vecs, the mirror Landau-Ginzburg theory has twisted superpotential
\eqn\lgsup{
\widetilde{W}=\sum_{a=1}^2\Sigma_a\left(\sum_{i=0}^7Q_{a,i}Y_i-t_a\right)+\sum_{i=0}^7e^{-Y_i}.
}
The period integral which computes BPS masses in the related compact theory is then given by\foot{In all manipulations, we suppress the explicit contours of integration, which are period-dependent.}
\eqn\lgper{
\Pi=\int d\Sigma_1 d\Sigma_2\Big(\prod_{i=0}^7dY_i\Big)\,\(5\Sigma_1-\Sigma_2\)\exp\left(-\widetilde{W}\right),
}
where the term $(5\Sigma_1-\Sigma_2)$ has been inserted to render the theory (partially) compact.  Manipulation of this period integral leads to a representation of the related B-model Calabi-Yau geometry as a hypersurface as follows:
\eqn\alg{\eqalign{
\Pi&=\int d\Sigma_1d\Sigma_2 \Big(\prod_{i=0}^7dY_i\Big)\;{\del\over\del Y_0}\left[\exp\left(-\sum_{a=1}^2\Sigma_a\Big(\sum_{i=0}^7Q_{a,i}Y_i-t_a\Big)\right)\right]\exp\left(-\sum_{i=0}^7e^{-Y_i}\right),\cr
&=\int d\Sigma_1d\Sigma_2 \Big(\prod_{i=0}^7dY_i \Big)\;e^{-Y_0}\exp\left(-\sum_{a=1}^2\Sigma_a\Big(\sum_{i=0}^7Q_{a,i}Y_i-t_a\Big)\right)\exp\left(-\sum_{i=0}^7e^{-Y_i}\right),\cr
&=\int\Big(\prod_{i=0}^7dY_i \Big)\;e^{-Y_0}\delta\Big(\sum_{i=0}^7Q_{1,i}Y_i-t_1\Big)\delta\Big(\sum_{i=0}^7Q_{2,i}Y_i-t_2\Big)\exp\left(-\sum_{i=0}^7e^{-Y_i}\right).
}}

At this point, there are two sets of manipulations which lead to different, but equivalent, representations of the B-model geometry.  In order to make contact with the results of section three, one may make the following change of variables,
\eqn\cov{\eqalign{
e^{-Y_0}&=P\cr
e^{-Y_i}&=e^{-t_1/5}P{z_1^5\over x_1x_2x_3x_4x_5}\qquad\qquad i=1,\ldots,5\cr
e^{-Y_6}&=Z,
}}
the first delta function is satisfied automatically. This change of variables is only one-to-one after dividing out by a $\IC^\star\times{\bf Z}_5^3$ action under which \cov\ is invariant.  Evaluating the second delta function to fix $Y_7$ as well, the periods become,
%
%
%
%
%
\eqn\lgperb{
\Pi=\int\Big(\prod_{i=1}^5{dx_i}\Big){dZ\over Z}\;\delta\Big(\cG(x_i)\Big)\exp\left(Z\Big(1+e^{-t_2-t_1/5}{x_1^5\over x_1x_2x_3x_4x_5}\Big)\right)
}
where 
\eqn\quint{
\cG(x_i)=x_1^5+x_2^5+x_3^5+x_4^5+x_5^5+\psi x_1x_2x_3x_4x_5,
}
and $\psi=\exp\(t_1/5\)$. Introducing new fields $\tilde{u}$ and $\tilde{v}$ and inserting the identity in the form,
$$
1=Z\int d\tilde{u}\,d\tilde{v}\,e^{\tilde{u}Z\tilde{v}},
$$
\lgperb\ becomes
\eqn\semifin{
\Pi=\int\Big(\prod_{i=1}^5dx_i\Big)du\,dv\,\delta\Big(\cG(x_i)\Big)\delta\left(\tilde{u}\tilde{v}+1+\phi{x_1^5\over x_1x_2x_3x_4x_5}\right).
}
A final coordinate redefinition,
\eqn\lastvar{\eqalign{
u&=x_1x_2x_3x_4x_5\tilde{u}\cr
v&=\tilde{v},
}}
leads to an expression which describes periods of the holomorphic four-form on a non-compact Calabi-Yau fourfold,
\eqn\lastper{
\Pi=\int\big(\prod_{i=1}^5dx_i\big)du\,dv\,\delta\big(\cG(x_i;\psi)\big)\,\delta\big(\cP(x_i;u,v;\phi)\big),
}
where
\eqn\hypers{
\cP(\phi)=uv+x_1^5+\phi\,x_1x_2x_3x_4x_5,
}
and $\phi=\exp\(t_2+t_1/5\)$. This precisely matches the fourfold which was obtained by T-duality considerations in section three.

Alternatively, starting with \alg, we can introduce variables as follows,
\eqn\vartwo{\eqalign{
e^{-Y_0}&=P\cr
e^{-Y_i}&=e^{-t_1/5}P{x_1^5\over x_1x_2x_3x_4x_5}\qquad\qquad i=1,\ldots,5\cr
e^{-Y_6}&=P{e^z\over x_1x_2x_3x_4x_5}.\cr
}}
Again, the first delta function is satisfied automatically, while the second delta function can be enforced to fix $Y_7$ in terms of the other variables.  The resulting periods are of the form
\eqn\finaleqs{
\Pi=\int\Big(\prod_{i=1}^5{dx_i}\Big)dz\,dP\;\exp\Big(P\;\cH(\psi,\phi)\Big),
}
where
\eqn\H{
\cH=x_1^5+x_2^5+x_3^5+x_4^5+x_5^5+\psi x_1x_2x_3x_4x_5+e^w(x_1^5+\phi x_1x_2x_3x_4x_5)
}
with $\psi$ and $\phi$ defined above. These periods are then given by integrals of the holomorphic four-form on the Calabi-Yau fourfold defined by $\cH=0$.  This formulation makes manifest the structure of the fourfold as a fibration of a Calabi-Yau threefold over a cylinder, as in \refs{\Mayrone,\AlimBX,\GrimmEF}.

A similar analysis to these can be carried out for the other toric examples.  However, the derivation in terms of T-duality is more general, since it also applies to cases which cannot be described within the ``toric brane'' framework.
\listrefs
\end